\documentclass[baaa]{baaa}

\usepackage[pdftex]{hyperref}
\usepackage{subfigure}
\usepackage{natbib}
\usepackage{helvet}
\usepackage[font=small]{caption}

\begin{document}


\journalvol{57}
\journalyear{2014}
\journaleditors{A.C. Rovero, C. Beaugé, L.J. Pellizza \& M. Lares}


\contriblanguage{0}


\contribtype{2}

\thematicarea{1}

\title{Angular momentum evolution for galaxies}
\subtitle{in a $\Lambda$ CDM scenario}


\titlerunning{Angular momentum evolution for galaxies}


\author{S.E. Pedrosa\inst{1}\& P.B. Tissera\inst{2}}
\authorrunning{Pedrosa et al.}
\contact{ACR: supe@iafe.uba.ar}

\institute{Instituto de Astronomía y Física del Espacio (CONICET-UBA) \and
  Departamento de Física, Universidad Andrés Bello)
}


\resumen{
Usando simulaciones hidrodinámicas cosmológicas estudiamos el contenido de momento angular de las galaxias simuladas en relación con su tipo morfológico. Encontramos que no solo la componente de disco sigue la relación teórica esperada, \citep{MMW98}, sino que también lo hace la componente esferoidal presentando un offset debido a su pérdida de momento angular, en acuerdo con \citep{fall2013}. También encontramos que el tamaño característico de la galaxia puede graficarse en una sola relación general aunque sean de tipos morfológicos distintos, como encontró \citep{kravtsov2013}.}

\abstract{
Using cosmological hydrodinamics simulations we study the angular momentum content of the simulated galaxies in relation with their morphological type. We found that not only the angular momentum of the disk component follow the expected theoretical relation, \citep{MMW98}, but also the spheroidal one, with a gap due to its lost of angular momentum, in agreement with \citep{fall2013}. We also found that the galaxy size can plot in one general relation, despite the morphological type, as found by \citep{kravtsov2013}.
}


\keywords{galaxies: evolution --- galaxies: formation --- galaxies: structure}


\maketitle

\section{Introduction}
\label{S_intro}

In the current cosmological paradigm, galaxy formation is a complex process where inflows, outlfows, interactions and mergers are common events. What determines the final morphology of galaxies is still a matter of debate. \citet{fall1980} provided a theoretical explanation to the formation of disc galaxies based on the hypothesis of specific angular momentum conservation. This model is able to explain the observed correlation between the angular momentum  of disc and the stellar mass \citep{fall1980}. 

Recently \citep{fall2013}, analysing obervational results, found that not only disk-dominated galaxies follow this relation but also elliptical ones follow a similar, nearly parallel, relation but offset to lower $j_{star}$ at each $M_{star}$. They found that observed galaxies of all morphological types lie along nearly parallel sequences with exponents $\alpha \sim 0.6$ in the $j_{star}-M_{star}$ diagram, with an offset between late and early type. 

On the other hand, \citep{kravtsov2013} found that characteristic size of stellar and gas distributions in galaxies scales approximately linearly with the virial radius that they derived using abundance matching approach. They found that the relation is in good agreement with expectations of the model of \citep{MMW98}. But they found that remarkably, this prediction works not only for late type disks, but also for early type galaxies, probably meaning that angular momentum plays a crucial role in determining the sizes of galaxies of all morphological types. 

In this work we analyse these scaling relations in simulated galaxies.
 
\section{Numerical Experiments}

We analysed cosmological simulations of a typical field region of the Universe consistent with the concordance model with $\Omega_{\Lambda}=0.7$, $\Omega_{\rm m}=0.3$, $\Omega_{b}=0.04$, a normalization of the power spectrum of $\sigma_{8}=0.9$ and $H_{0}= 100 h \ {\rm km} \ {\rm s}^{-1}\ {\rm Mpc}^{-1}$, with $h=0.7$. 
The simulations were performed by using the code {\small GADGET-3}, an update of {\small GADGET-2 } \citep{springel2003, springel2005}, optimized for massive parallel simulations of highly inhomogeneus systems. This version of {\small GADGET-3} includes treatments for metal-dependent radiative cooling, stochastic star formation (SF), chemical enrichment, and the multiphase model for the interestellar medium (ISM) and the Supernova (SN) feedback scheme of \citet{scan2005,scan2006}. 
This SN feedback model is able to successfully trigger galactic mass-loaded winds without introducing mass-scale parameters fact which makes it specially suitable for the study of galaxy formation in cosmological context. 

For this study we use one of the simulations run as part of the Fenix Project. The Fenix simulations share the same initial conditions but the parameters that regulates the star formation and SN feedback are different (Tissera et al. in preparation). In order to clasify our simulated galaxies into early and late type we adopt the criteria of \citet{tissera2012}.

\section{Results and Discussion}

We study the angular momentum component of our simulated galaxies. We used the classical notation of \citet{MMW98} in term of the fraction of the galaxy angular momentum with respect to the dark matter halo one. So we estimated the ratio between  the specific angular momentum of stars in the discs  within the galactic radius, $r_{\rm gal}$, and the corresponding of the dark matter haloes within the virial one $J_{\rm D} /J_{\rm H} = j_{\rm d}$ as a function of the ratio between the corresponding masses  $M_{\rm D} /M_{\rm H} = m_{\rm d}$. We calculted this fraction for both, the disc and spheroidal components. We found a clear correlation with the angular momentum content of the dark matter halo, not only for the disk component but also for the central spheroid, as we have already found in \citet{Pedrosa2014} for a slight different tune of the feedback parameters. 

As assumed in many semianalitical models, there is a one to one relation between $j_{d}$ and $m_{d}$, for the disc component. Remarkably we also found that the spheroidal component share the same relation, but with an offset to lower values. It is clear that the last one is made of low angular momentum material.
The one-to-one correlation is based on the fact that all the material in the system experiences the same external torques before separating into two distinct components. Although the angular momentum content of both components can be affected by mergers, the SN feedback and star formation, still the correlation is conserved.
The lost of angular momentum of the spheroidal component is directed related with the fact that in this case a rotationally soported structure wasn't able to form or may be it forms at higher redshifts but then, due to mergers and interactions, it loses part of its angular momentum.

This correlation can be linked with the morphological type of the galaxy. In agreement with \citet{fall2013}, we found that the specific angular momentum content of both disk and spheroidals follow a parallel sequence. We found that higher D/T ratios are related with higher contents of specific angular momentum, as expected. It also present a good agreement with the observational trend found by \citet{fall2013}. 

On the other hand, the size of galaxies is related to the originally rotationally supported gas disk, which size depends on the angular momentum content of the gas. \citet{kravtsov2013} investigated this relation. He found that characteristic size of stellar and gas distributions in galaxies spanning several orders of magnitude in stellar mass scales approximately linearly with a fraction of the virial radius derived using abundance matching technique. We have calculated the half mass radius of the stellar component of our simulated galaxies and found that for our simulated galaxies a similar relation can also be constructed. Both, late and early type galaxies can be plotted in the same linear relation. But with a different factor multiplying the virial radius.

\subsection{Acknowledgement}
\label{thanks}

\begin{acknowledgement}
This work was partially supported by PICT 2011-0959 from ANPCyT, PIP 2009-0305 and PIP 2012-0396 awarded by CONICET from Argentina. PBT thanks support from the Millennium Institute of Astrophysics (MAS) and the Regular Grant UNAB 2014. Simulations were run in the Fenix cluster of the Numerical Astrophysics Group at the Institute for Astronomy and Space Physics.
\end{acknowledgement}


\bibliographystyle{baaa}
\small
\bibliography{fenix}

\begin{thebibliography}{}

\bibitem[\protect\citeauthoryear{{Fall} \& {Efstathiou}}{{Fall} \&
  {Efstathiou}}{1980}]{fall1980}
{Fall} S.~M.,  {Efstathiou} G.,  1980, \mnras, 193, 189

\bibitem[\protect\citeauthoryear{{Fall} \& {Romanowsky}}{{Fall} \&
  {Romanowsky}}{2013}]{fall2013}
{Fall} S.~M.,  {Romanowsky} A.~J.,  2013, \apjl, 769, L26

\bibitem[\protect\citeauthoryear{{Kravtsov}}{{Kravtsov}}{2013}]{kravtsov2013}
{Kravtsov} A.~V.,  2013, \apjl, 764, L31

\bibitem[\protect\citeauthoryear{{Mo}, {Mao} \& {White}}{{Mo}
  et~al.}{1998}]{MMW98}
{Mo} H.~J.,  {Mao} S.,    {White} S.~D.~M.,  1998, \mnras, 295, 319

\bibitem[\protect\citeauthoryear{{Pedrosa}, {Tissera} \& {De Rossi}}{{Pedrosa}
  et~al.}{2014}]{Pedrosa2014}
{Pedrosa} S.~E.,  {Tissera} P.~B.,    {De Rossi} M.~E.,  2014, \aap, 567, A47

\bibitem[\protect\citeauthoryear{{Scannapieco}, {Tissera}, {White} \&
  {Springel}}{{Scannapieco} et~al.}{2005}]{scan2005}
{Scannapieco} C.,    et~al., 2005, \mnras, 364, 552

\bibitem[\protect\citeauthoryear{{Scannapieco}, {Tissera}, {White} \&
  {Springel}}{{Scannapieco} et~al.}{2006}]{scan2006}
{Scannapieco} C.,    et~al., 2006, \mnras, 371, 1125

\bibitem[\protect\citeauthoryear{{Springel}}{{Springel}}{2005}]{springel2005}
{Springel} V.,  2005, \mnras, 364, 1105

\bibitem[\protect\citeauthoryear{{Springel} \& {Hernquist}}{{Springel} \&
  {Hernquist}}{2003}]{springel2003}
{Springel} V.,  {Hernquist} L.,  2003, \mnras, 339, 289

\bibitem[\protect\citeauthoryear{{Tissera}, {White} \& {Scannapieco}}{{Tissera}
  et~al.}{2012}]{tissera2012}
{Tissera} P.~B.,  {White} S.~D.~M.,    {Scannapieco} C.,  2012, \mnras, 420,
  255

\end{thebibliography}
 
\end{document}